\documentstyle[preprint,eqsecnum,aps]{revtex}
\begin{document}
\draft
\def \beq{\begin{equation}}
\def \eeq{\end{equation}}
\def \beqarr{\begin{eqnarray}}
\def \eeqarr{\end{eqnarray}}
\def\bc{\begin{center}}
\def\ec{\end{center}}
\def \ca{\cos \alpha }
\def \ct{\cos \theta_{u}}
\def \st{\sin \theta_{u}}
\bc 
\title  {Quantum Hall Solitons with Intertwined Spin and Pseudospin  
 at $\nu \ = \ 1$} 
\author{Sankalpa Ghosh and R. Rajaraman\cite{byline1}}

\address{School of Physical Sciences \\
Jawaharlal Nehru University\\ New Delhi 110067, \ INDIA\\ }

\maketitle

\ec
\begin{abstract}
In this paper we study in detail different types of topological
solitons which are possible in bilayer quantum Hall systems at filling 
fraction $\nu =1$ when spin degrees of freedom are included. Starting from a
microscopic Hamiltonian we  derive an effective energy functional
for studying such excitations. The gauge invariance and $CP^{3}$ character
of this energy fuctional and their consequences are examined.
 Then we identify permissible classes of 
finite energy solutions  which are topologically non-trivial. 
 We also numerically evaulate 
a representative solution in which a pseudospin (layer degrees of freedom)
bimeron in a given spin component is intertwined with spin-skyrmions 
in each layer ,  
and and discuss whether it is energetically favoured as
the lowest lying excitation in such system with some numerical results.
\end{abstract}
\vskip 16pt

\section{Introduction}
Systems which permit topological excitations, i.e. where  field
configurations can be classifield by  homotopy sectors
characterised typically by some winding number, have been
studied in a general sense in mathematical physics for a long
time.  That such interesting possibilities can actually arise
and play a significant role in the Quantum Hall physics was
demonstrated in the work of  of Sondhi {\it et al} \cite
{Sondhi}. They showed that for example , in a single layer Hall
liquid at filling factor $\nu = 1$, the lowest energy
excitations in spin, for low Zeeman coupling  are the so-called
Skyrmions and not single spin flips.  These Skyrmions are
topological excitations in the spin texture, in which the spin
starts being, say, "up"  at the origin and as you go outwards,
starts tilting down in a flared manner to become asymptotically
"down" spin at large distances.  Subsequently experimental
support for the existence of such excitations was also
discovered in NMR measurements \cite {Barrett}.

Meanwhile  Quantum Hall phenomena have also been studied in
double layer systems \cite{Eisen}, \cite{Tapas}. The double well
Hall plateaux at unit filling can be understood by associating with
each electron a "pseudospin" in addition to its   lowest Landau
level (LLL) orbital wavefunction \cite{GirvMac}, \cite{Moon}.
The up and down components of this pseudospinor give the
probability amplitudes for the electron being in the upper and
lower layer respectively. The ground state of the $\nu = 1 $
double layer system, known to be a quantum Hall state with a
Hall conductivity plateaux is a pseudospin ferromagnet, with the
pseudospin aligned in the x-direction.  This is a very
remarkable phenomenon in that it amounts to interlayer coherence
between the electrons in the two layers.  This pseudospin degree
of freedom is in addition to physical spin.  To start with, in
analysing double layer phenomena, the spin degrees of freedom
are suppresed for simplicity. Even then one can still consider
excitations in the pseudospin. Inspired by the presence of
Skyrmions in spin, people have also considered the possibility
of topological excitations in pseudospin.   Such pseudospin
textures called "merons" and "bimerons" have been suggested as
possible low lying excitations of double layer systems
\cite{GirvMac}. The homotopy group $\pi_{2}[S_{2}]$ and its
winding number are identical for spin and pseudospin since
mathematically pseudospin is identical to spin, both being SU(2)
spinor fields on a plane.  The change in  terminology  from
Skyrmions to bimerons does not indicate any topological
difference between the two in going from spin to pseudospin
excitations but only differences in their detailed profiles.
This difference in turn happens because of the difference in the
energetics of  spin and pseudospin and correspondingly , their
asymptotic direction.  Meron  excitations , if present in double
layers, can give rise to a Kosterlitz-Thouless (KT) \cite{KT}
transition which may be enable them also to be  experimentally
observable.
    
Clearly there are prospects of even more esoteric excitations
when both  spin and pseudospin degrees of freedom are considered
simultaneously. That is the theme of this present work. We will
continue to study the  unit filling factor ($ \nu = 1 )$ case.
There has already been some discussion of the combined
spin-pseudospin $\nu = 1 $ double layer system  \cite{Arov},
\cite{Ezawa}. Our work  discusses diferent aspects of the
problem than these studies. We analyse in substantive detail
intertwined spin-pseudospin topological excitations of this
system . When both spin and pseudospin are active degrees of
freedom , these are together described by a 4-component object.
This 4-component object has been referred to as a $CP_{3}$
spinor in the literature \cite{Ezawa}.  That is correct, but
needs to be justified. A theory does not become  a $CP_{3}$
theory just because its field is a normalised 4-component
object. The system must enjoy a U(1) gauge invariance, which is
what makes the spinors span a projective space, implied in the
acronym CP. Without that gauge invariance the beautiful results
on $CP_{N}$ in the literature \cite{CP}, \cite{Raj} cannot be
borrowed.
 
 So we begin in sec. II  by showing, starting from the basic
microscopic theory of the $\nu = 1 $ system that  in the
effective LLL theory for the  spin-pseudospin texture such
gauge invariance $\underline{is}$ there. This is a
straightforward derivation following the procedure developed by
Moon et al \cite{Moon}.  In fact we find that in the limit where
the layer-separation $d$ vanishes, the Coulomb interaction
energy is precisely the protoype $CP_{3}$ Euclidean action used
in the pioneering papers  on that topic \cite{CP}, for which
exact topological solutions are known in terms of analytic
functions.  Of course, when $d\neq 0$, the energy functional is
more complicated and these analytical solutions do not hold.
But the theory is still a $CP_{3}$ theory, and the homotopy
classification of the solutions still holds. Only the solutions
themselves have to be calculated numerically.

 A topologically non-trivial $CP_{3}$ solution will generally
involve an intertwined texture in the physical spin in each
layer as well as in the pseudospin of each physical spin
projection.  One can ask whether such solutions can be
legitimately interpreted as containing, as subsystems,
spin-Skyrmions in either or  both of the layers, possibly
intertwined with a pseudospin meron or bimeron . If so, then
such possibilities of containing several  topological entities
as subsystems has to be made compatible with the fact that any
finite energy $CP_{3}$ texture carries altogether
$\underline{only  \ one}$ topological winding number. We study
all these questions in sec.III and find that there are certain
restrictions on the types of solutions permitted. We show that
the individual layers of a double layer system cannot accomodate
all possible spin structures one may find in a pair of unrelated
single layers. The spin winding numbers in the two layers are
related to one another and to the  pseudospin winding number.
 
 Consistent with these restrictions, we then pick in sec IV a
representative ansatz which can be viewed as a spin-Skyrmion
intertwined with a pseudospin bimeron. We then numerically
evaluate such a solution by solving the coupled nonlinear
partial differential equations that arise from extremising the
texture energy functional.  In earlier work
\cite{Ghosh1},\cite{Ghosh2} we had studied in some detail both
meron and bimeron excitations in pseudospin for double layer
systems, with the spin degree of freedom suppresed. The present
calculation is a more complicated version with  $CP_{3}$
spinors, but is done by similar numerical techniques. We present
the spin and pseudospin proflies of our intertwined solutions
for different values of interlayer separation.
 
 We also estimate the interaction energy of these solutions for
some typical sets of values of system parameters.  We discuss
the dependence of this energy on the
separation between the two meron centers. We find , as expected, 
that if only the gradient and 
capacitance energies are considered ,their minimisation will drive
the textures towards zero size. Therefore we also calculate the 
topological charge dependent Coulomb energy of
our solutions which, being repulsive ,  should drive the merons farther 
apart, offsetting the above tendency towards zero size.
 Then we extremise the total energy so obtained and find that it does show a
 minimum at some optimal meron separation, for each value of layer 
 separation. 
 
 We also find that these
energies are approximately of the same order as those of purely
spin Skyrmions of the single layer system. We make qualitative speculations
 on whether or not our spin-pseudospin intertwined solitons can be
energetically favoured over solitons purely in spin or
pseadospin, or over simple spin-flips.  
 
\section{Texture Energy and Its Gauge Invariance}
In a double layer quantum Hall system with both spin and pseudospin 
 degrees of freedom present,  an electron 
 will carry, apart from its coordinate wavefunction $\phi_{X}({\vec r})$,
a 4-component normalised spinor whose components 
in general may vary with the orbital
quantum number $X$. For any given $X$, this spinor can be denoted by

\beq    a_{\sigma}(X) \ \ \ \ \ =
 \pmatrix{  
               a_{1}(X) \cr
         a_{2}(X) \cr
	a_{3}(X) \cr
            a_{4}(X)   \cr} \label{spinor} \eeq
where the spin-pseudospin index $\sigma = 1,2,3,4$ corresponds
to amplitudes that the electron is in the upper-layer up-spin,
upper-layer down- spin, lower layer up-spin and lower -layer
down-spin states respectively. It will henceforth be understood that the
spinor is normalised, i.e.$ \sum_{\sigma} \mid a_{\sigma}(X) \mid^2 = 1 $
for each $X$
In the literature, this $a_{\sigma}$ has sometimes been referred
to as a $CP_3$ spinor (see for instance Ezawa \cite{Ezawa}).
 That is correct, but requires a little
justification.  In a $CP_3$ theory, the spinor must not only be
normalised 4-component object, but be defined only modulo a local gauge
transformation common to
all four components. This in turn requires that the Euclidean
action or static energy functional of the spinor field 
enjoy a corresponding gauge
invariance. In this section we will verify all this. We will
also see that the nature of the gauge symmetry is different for
a double layer system than for a pair of isolated single layers.
 This, as we shall see, has the important consequence of
\underline{prohibiting} certain topological spin excitations in the double
layer system which would have been present in the individual layers 
had they been far apart. In this way, along with 
establishing the $CP_3$ nature of the system we
will also  identify permissible types of excitations where
the spin and pseudospin are nontrivially intertwined, some of which we 
numerically evaluate in later sections.

Let us start by deriving the energy functional of any spin-pseudospin
texture from the microscopic Hamiltonian. This is just a straightforward
generalisation of the procedure already in the literature for the simpler
case of a spinless bilayer problem \cite{Moon}. Therefore we need to present
only the essential equations needed for completeness and understandability.
 We take the microscopic Hamiltonian to be 
\beq H \ = \ H_{K} + H_{1} +  H_{C} \eeq
Here
\beq H_{K} \ = \ \frac{1}{2m} \sum_{\sigma=1}^{4} \int d\vec r   
\psi^{\dag}_{\sigma} \ D^{2} \ \psi_{\sigma} \label {HK3} \eeq  
is the kinetic energy in the presence of the magnetic field. We will be
working at $\nu = 1$ in the lowest Landau level (LLL) approximation.
Corespondingly, the operator
$\psi_{\sigma}(\vec r)$ is the LLL-projected electron field operator
 expanded in terms of Lowest Landau Level orbitals as
\beq  \psi_{\sigma}(\vec r) =  \sum_{X = 1}^{N}
       \phi_{X} (\vec r) \ C_{\sigma X}  \eeq
with $\phi_{X} (\vec r)$ being a LLL orbital, say, in the Landau gauge
 with $X$ as its  guiding center.       

The second term in the Hamiltonian is the one
body term representing the Zeeman and interlayer tunnelling energies.
\beq H_{1} \ = \   \ \sum_{\sigma,\delta } \int d\vec r   
\psi^{\dag}_{\sigma}(\vec r) \ \big( \tilde{g} \hat{\sigma}_{z} \ - \ t 
\hat{\tau}_{x} \big)
_{\sigma \delta} \ \psi_{\delta}(\vec r) \eeq
where $\hat{\sigma}_{z}$ and $\hat{\tau}_{x}$ are  spin and pseudospin 
matrices suitably generalised as $4 \times 4$ matrices on the outer
product space of spin and pseudospin.

The third term in the Hamiltonian is the Coulomb term :
\beq H_{C} \ = \ \frac{1}{2} \sum_{\sigma_{1},\sigma_{2} =1}^{4} 
\int d \vec r_{1}
d \vec r_{2}\psi^{\dag}_{\sigma_{1}}(\vec r_{1})\psi^{\dag}_{\sigma_{2}}
(\vec r_{2})V^{\sigma_{1} \sigma_{2}}
(\vec r_{1} -\vec r_{2})\psi_{\sigma_{2}}(\vec r_{2})\psi_
{\sigma_{1}}(\vec r_{1}) \label{Hint} \eeq
 In the above , the 
Coulomb potential $ V^{\sigma_{1} \sigma_{2}}$ depends on 
whether the particles
are in the same layer or different layers
\beqarr V^{\sigma_{1} \sigma_{2}} \ &=& v^{s} \ \equiv 
{ e^2 \over \epsilon  \ r_{12}} \ \ \ \ 
 .... \ \ {\sigma_{1} \ , \   \sigma_{2} } \ in  \ same  \ layer, \nonumber \\
\ V^{ \sigma_{1} \sigma_{2}} \ &=& v^{d} \ \equiv 
{ e^2 \over \epsilon \sqrt{r_{12}^{2} + d^{2}}} \ \ \ \ 
 \ .... \  {\sigma_{1} \ , \  \sigma_{2}}  \ in  \ different  \ layers \ \eeqarr
where $d$ is the interlayer distance.
To obtain the energy of an arbitrary spin-pseudospin texture, we adopt the 
strategy followed in the work of Moon et al \cite{Moon}. We first consider the
ansatz  state
\beq \mid \Psi \rangle \ = \ \prod_{X} \ [ \sum_{\sigma} \ 
C^{\dagger}_{\sigma X}  a_{\sigma}(X) ] \ \mid 0 \rangle  \label{state} \eeq
where $\mid 0 \rangle$ is the vacuum (no electron) state , $X$ stands
for Landau gauge orbitals  and $a_{\sigma}(X)$ is an orbital
dependent 4-spinor as in eq.(2.1).  In the high-B limit  each
Landau gauge orbital density is uniform along the y-axis with 
support on a thin line localised around some value of x.
Further these states are closely spaced along the x-direction.
Using this feature, we will later on replace the orbital label $X$ by
the x-coordinate itself. In that case the above texture    $a_{\sigma}(X)$
depends only on the x-coordinate and not on y, and therefore carries
zero topological number density (see eq.( 3.14) below). 
Nevertheless we will use this ansatz to calculate its energy functional, 
and then later that energy functional to the more general and 
topologically non-trivial textures by invoking isotropy of the system in the 
x-y plane. This was the exactly the strategy used in ref(\cite{Moon}).
We will calculate the energy functional of the spin-pseudospin texture (2.8)
by taking the mean value of the second quantised Hamiltonian  in 
that state .

At unit filling $\nu=1$, and in the space of LLL orbitals 
the kinetic term $H_K$  is just a constant equal to 
$(N/2) \hbar \omega $ ,the energy of the filled
LLL band. This constant will henceforth be neglected. 

The Zeeman and tunnelling one-body energies yield
\beqarr E_{1} [ a_{\sigma}(X) ] \ &=& \  \ \sum_{X} \bigg[ \tilde{g}  \ 
\bigg(|a_{1}(X)|^{2} - |a_{2}(X)|^{2} +
 |a_{3}(X)|^{2} - |a_{4}(X)|^{2} \bigg)  \nonumber \\
 &-& \ t \bigg(a_{1}(X)a_{3}^{*}(X) + 
 a_{2}(X)a_{4}^{*}(X) \ +  \ c.c. \ \bigg) \bigg] \label{1body} \eeqarr
  
The expectation value of the Coulomb interaction Hamiltonian can be
convenientlly written in terms of  the following
spinorial bilinears for the upper(u) and lower(l) layers 
\beqarr  F_{u}(X) & = & |a_{1}(X)|^{2} + |a_{2}(X)|^{2}  \label{F1} \\
         F_{l}(X) & = & |a_{3}(X)|^{2}  + |a_{4}(X)|^{2}   \label{F2} \\
 G_{u}(X_{1},X_{2})    & = & \sum_{i=1,2} a^{i}(X_{1})a^{i \ast}(X_{2})
\label{G1} \\
    G_{l}(X_{1},X_{2}) & = & \sum_{i=3,4} a^{i}(X_{1})a^{i \ast}(X_{2}) 
    \label{G2} \eeqarr
On inserting $H_C$ from (\ref{Hint}) and the state
 $\Psi$ from (\ref{state}) straightforward algebra then gives  
us the Coulomb  
 energy in terms of the spinors $a_{\sigma}$.
 \beq E_{C} [ a_{\sigma}(X) ] \equiv 
 \langle \Psi \mid H_{C} \mid \Psi \rangle
= \langle H_{C} \rangle_{direct} - \langle H_{C} \rangle_{exchange}
\label{energy} \eeq
with 
\beq \langle H_{C} \rangle_{direct}  \ \ 
    =  \ \  \frac{1}{2} \sum_{X_{1},X_{2}} \ \bigg[
D^{s} 
 + (D^{d} - D^{s} )
( F_{u}(X_{1})F_{l}(X_{2}) +  F_{l}(X_{1})F_{u}(X_{2}) ) 
\bigg] \label{dir} \eeq
and
\beq \langle H_{C} \rangle_{exchange} =  
\frac{1}{2} \sum_{X_{1},X_{2}}  \bigg[ \ E^{s}(|G_{u}|^{2} +
|G_{l}|^{2}) + E^{d}(G_{u}^{\ast}G_{l} + G_{u} G_{l}^{\ast} ) 
\bigg] \label{ex} \eeq

Here
 \beqarr D^{s,d} \ (X_{2}-X_{1}) \ &=&  \  V^{s,d}_{X_1 , X_2 , X_1 , X_2} 
 \nonumber \\            
 E^{s,d } (X_{2}-X_{1}) \ &=&  \  V^{s,d}_{X_2 , X_1,  X_1 , X_2}  \eeqarr
with
\beq V^{s d}
_{X_{1},X_{2},X_{3},X_{4}} \ = \ \int d\vec r_{1} d\vec r_{2}
V^{s,d}
(\vec r_{1} - \vec r_{2}) \times \phi^{\ast}_{X_{1}}(\vec r_{1})
\phi^{\ast}_{X_{2}}(\vec r_{2}) \phi_{X_{3}}(\vec r_{1})
\phi_{X_{4}}(\vec r_{2}) \eeq
These direct and exchange Coulomb interaction matrix elements $D^{s,d}$
 and $E^{s,d}$ between two
electrons in LLL orbitals $X_1$ and $X_2$, in the same(s) or different(d) 
layers, are 
exactly the same as were used in the spinless double layer problem by 
Moon {\sl et al} \cite{Moon}. However, the inclusion 
of the physical spin degrees
of freedom is reflected in the energy expressions in \ref{dir}
 and \ref{ex} , which involve all four components of the spin-pseudospin
 multiplet $a_{\sigma}$.

Adding the contributions in eq.(\ref{energy}) and eq.(\ref{1body}) 
we get the total energy expectation value
\beq E [ a_{\sigma}(X) ] \ = \ E_{1} [ a_{\sigma}(X) ]  \ +  \ 
E_{C} [ a_{\sigma}(X) ] \label{Etot} \eeq
In the Hartre-Fock approximation, this energy expectation value 
$E [ a_{\sigma}(X) ]$ in eq.(\ref{energy}) will be minimised to
get the ground state and excited state spin-pseudospin textures.

But, let us first examine the  gauge invariance of the
 energy functional $E [ a_{\sigma}(X) ]$. Consider the transformation
\beqarr  a_{\sigma}(X) \ &\rightarrow \ e^{i \Lambda_{u}(X)}  
a_{\sigma}(X)  \ \ \ \ \ for \ \  \sigma =1,2 \nonumber \\
a_{\sigma}(X) \ &\rightarrow \ e^{i \Lambda_{l}(X)}  
a_{\sigma}(X)  \ \ \ \ \ for  \ \ \sigma =3,4 \label{gtr} \eeqarr 
Notice that we have used  different  phases $\Lambda_{u}(X)$  
and $\Lambda_{l}(X)$ for the  upper and lower layer components
 respectively. This is a $U(1) \times U(1)$ transformation. These 
 phases can also vary with the orbital index $X$. 
[ Note : $X$  is not the space coordinate. But, following
accepted approximations  (see ref \cite{Moon}) eventually the
sum over the orbital index $X$ will  be converted into an
integral over space coordinate, invoking the fact that for large magnetic
fields, each LLL orbital wavefunction is highly localised. 
Hence  the above $X$ dependent
transformation corresponds to spatially local gauge
transformations.] 

Under these local $U(1) \times U(1)$ transformations, the one-body
Zeeman energy in (\ref{1body}) and the direct part of the Coulomb energy
(\ref{dir}) are trivially invariant since they involve only the 
squared-modulus
of $a_{\sigma}(X)$ . So is the first part (proportional to $E^{s}$ ) 
of the exchange Coulomb energy (\ref{ex}) . But the tunnelling energy in
(\ref{1body}) and the second piece of the exchange energy (\ref{ex}),
which involves exchange Coulomb interaction $E^{d}$ between different layers
 are invariant only if 
 \beq \Lambda_{u}(X) \ = \ \Lambda_{l}(X) \ = \ \Lambda (X) \eeq 
 Thus the full energy of the double layer system enjoys only a 
 U(1) subgroup of $U(1) \times U(1)$ defined in (\ref{gtr}) -- 
 a subgroup where all 4 components of $a_{\sigma}$ are transformed by the
 $\it{same}$ phase.  This is the U(1) gauge invariance 
 modulo which our $CP_3$ spinors are defined.
 
 Consider,  however, what would happen if we had
 very  widely separated $\nu = 1$ layers 
 (the separation $d \rightarrow \infty$).
Then each can have its own 2-component spin texture 
 described by a $CP_1$ system
 (equivalent to a non-linear O(3) sigma model) with its own U(1)
  gauge symmetry (see \cite{GirvMac} and \cite {Raj}). 
  The well separated pair of layers should enjoy  $U(1) \times U(1)$
gauge  symmetry. Our derivation shows a similar effect. When
  $d \rightarrow \infty$,
  both the tunnelling parameter $t$ in (\ref{1body}) and the interlayer
 Coulomb potential $v^{d}$ involved in (\ref{ex})
 would vanish and the full  $U(1) \times U(1)$ gauge invariance would 
 indeed be restored. 
   We will see later that this reduced gauge symmetry of a double-layer system
at finite separation  has consequences in terms of what types of finite
  energy excitations are permitted in it  as compared
  to a pair of isolated single layers.

\section{Gradient Expansion and the $CP_3$ Field Theory}

To rewrite the energy expression \ref{Etot} in a continuum field theory 
language, we proceed following Moon {\sl et al} \cite{Moon} and convert 
sums over the LLL label $X$ into an integral over  space. 
Clearly the one body energy (\ref{1body}), which involves only a single sum
over the index $X$,  will become a  local term, i.e. a spatial integral over 
the 1-body energy density. But the Coulomb term (\ref{energy}) containing
 a double sum over
$X_{1}$ and $X_{2}$ will become a non-local term involving a double integral
over some coordinates $x_{1}$ and  $x_{2}$. For long wavelength
excitations one then makes the usual gradient expansion . Expand the spinor 
for $X_2$ as 
\beq a_{\sigma}(X_{2}) \ = \  a_{\sigma}(X_{1}) \ + \ ((X_{2}-X_{1})
{\partial \over \partial X_{1}} a_{\sigma}(X_{1}) \ + \ ..\label{grad}... \eeq
Up till now we found the energy of textures which were y-independent.

Now we will invoke the isotropy of the basic system in the x-y plane
and generalise this expression for arbitrary textures by 
This is done by  
making the replacement
\beq \sum_{X} \rightarrow {1 \over 2\pi l^{2}} \int d^{2}r \eeq
and by replacing x-derivatives by gradients.
 Insert the above expansion (\ref{grad}) into the Coulomb energy
 expressions (2.13 - 2.16). 
 Keep  terms only upto order $\partial_{X_{1}}^2$  and replace the sum 
$\sum_{X_{1}}$ by an integral over space as indicated. (These 
steps are given in the Moon et al work \cite{Moon} for  the simpler
spinless double layer case). The result, for our problem , is the following
local expression for the total energy (\ref{Etot}), with overall constants 
subtracted out :
    
\beqarr  E [a_{\sigma}]
& = &
  \frac{1}{2 \pi l^{2}} \ \int d{\bf r} \bigg[  \tilde{g}  \ 
\bigg(|a_{1}|^{2} - |a_{2}|^{2} +
 |a_{3}|^{2} - |a_{4}|^{2} \bigg) 
 - \ t \bigg(a_{1} a_{3}^{*} + 
 a_{2} a_{4}^{*} \ +  \ h.c. \ \bigg) \bigg] \nonumber \\
  &+& \beta_{m} 
\int d{\bf r}( F_{u}(\vec r) - F_{l}( \vec r))^{2} \nonumber \\
 &+& 
2\rho^s  \int d{\bf r} \bigg[
\sum_{i=1,4}(\partial_{\mu}a^{i\ast}(\vec r)\partial^{\mu}a^{i}(\vec r))
+(\sum_{i=1,4}a^{i \ast}(\vec r)\partial_{\mu}a^{i}(\vec r)^{2} 
\bigg] \nonumber \\             
&+& (\rho^{d} - \rho^{s}) 
 \int d{\bf r}
 \bigg[a^{1}a^{3 \ast} \vec \nabla^{2} (a^{3}a^{1 \ast})
+ a^{1}a^{4 \ast} \vec \nabla^{2} (a^{4}a^{1 \ast}) \nonumber \\
                                                        &     & \mbox{}   +
a^{2}a^{3 \ast} \vec \nabla^{2} (a^{3}a^{2 \ast}) +
a^{2}a^{4 \ast} \vec \nabla^{2} (a^{4}a^{2 \ast})  \  +  \ h. c. \bigg]
\label{INTE} \eeqarr

where the constants appearing above are defined by
\beq  \beta_{m} \ = \  \frac{1}{4} \sum _{(X_{2} - X_{1})} \bigg(
 (E^{d}(X_{2} - X_{1}) - E^{s}(X_{2} - X_{1}))
- (D^{d}(X_{2} - X_{1}) -D^{s}(X_{2} - X_{1})) \bigg) \eeq
\beq \rho^{s} \ = \frac{1}{2}
\ \sum_{ (X_{2} -X_{1})}
\frac{(X_{2} -X_{1})^{2}}{2}  \ E^{s}(X_{2} - X_{1})\eeq
\beq \rho^{d} \ = \frac{1}{2}
\ \sum_{(X_{2} -X_{1})} 
\frac{(X_{2} -X_{1})^{2}}{2}  \ E^{d}(X_{2} - X_{1})\eeq
These constants are again the same as given by Moon
{\it et al} \cite{Moon} in the spinless double layer problem.
The term involving $\beta$ 
represents the "capacitance energy" of the double layer system.
It is proportional to the square of $  F_{u}(\vec r) - F_{l}( \vec r)$,
which gives the difference in charge density between the
two layers. The constants $\rho^{s}$ and $\rho^d$ represent 
spin-pseudospin stiffness coming from intralayer and interlayer 
Coulomb interaction respectively.

This energy functional \ref{INTE} will act as the effective classical 
Hamiltonian to be minimised to find different textured solutions. 
The ground state will correspond to a spatially uniform texture, and 
so can be obtained by minimising the gradient-free terms in \ref{INTE}.
This is acheived  by the spinor $a_{\sigma}(X) = 
{1 \over \sqrt{2}} (0, 1, 0, 1) $. The one-body Zeeman and tunnelling 
energies are clearly minimised
by this choice since the  spin is
polarised "down" in both layers and the psuedospin is along the x-direction,
i.e. a layer-symmetric state. 
This choice also minimises the capacitance energy since it has 
 equal occupancy in the two layers $F_{u} = F_{l} = {1 \over 2}$.
 
 Moving on to excited states with non-trivial textures,
  these are obtained by extremising the full energy functional 
   (\ref{INTE}). Note that  (\ref{INTE}) including its gradient terms is 
   still gauge invariant under the local U(1) transformation  mentioned
    earlier,
\beq a_{\sigma}(X) \ \rightarrow \ e^{i \Lambda(X)}  
a_{\sigma}(X) \label{GT}\eeq
 so that this is still a $CP_3$ theory.
In fact the term proportional to the isotropic spin-pseudospin stiffness
$\rho^{s}$ 
, namely ,
\beq E_{CP}  \ \equiv \ 2\rho^{s} \int d{\bf r} \big{[}
\sum_{i=1,4}(\partial_{\mu}a^{i\ast}(\vec r)\partial^{\mu}a^{i}(\vec r))
+(\sum_{i=1,4}a^{i \ast}(\vec r)\partial_{\mu}a^{i}(\vec r)^{2} 
\big{]}  \label{proto} \eeq
is the Euclidean action for the prototype minimal $CP_3$  theory
\cite{CP}. Indeed, in  the limit where the layer separation $d$ is zero,
 this $E_{CP}$ will
be the only surviving term from the Coulomb energy in \ref{INTE} since 
the interlayer and interlayer Coulomb potentials 
will become equal ( $(v^{s} = v^{d})$) and 
hence both $\beta$ and $\rho^{s}- \rho^{d}$ will vanish.

The properties of this prototype $CP_3$ system and its topological
solitons are well known \cite{CP},
\cite{Raj}. Let us briefly recall those salient features which will be of 
relevance to us.
 Define a gauge field $A_{\mu}$ as follows.
\beq A_{\mu}  \equiv  i \sum_{\sigma} [a_{\sigma}^{*} \partial_{\mu} 
a_{\sigma}  ] \eeq
Clearly under the  gauge transformation \ref{GT} ,
\beq A_{\mu} \  \rightarrow A_{\mu} \ - \  \partial_{\mu} \Lambda \eeq
The energy $E_{CP}$ can then be written in a manifestly 
gauge invariant manner as 
\beq E_{CP}  \ \equiv \ 2\rho^{s} \int d{\bf r} \big{[}
\sum_{\sigma=1}^{4} \ \sum_{\mu=1}^{2} \ \mid D_{\mu}a^{\sigma} (\vec r) 
\mid^{2} \eeq where 
$D_{\mu} = \partial_{\mu} + i A_{\mu}$ is the covariant derivative of the
U(1) gauge transformation.
Then any finite energy field must obey, as ${\vec r} \rightarrow \infty$,
the boundary condition
\beq D_{\mu}a_{\sigma} = (\partial_{\mu} + i A_{\mu})a_{\sigma} = 0 \eeq
Since $A_{\mu}$ is independent of the spinor index $\sigma$, this
 implies ( see \cite{Raj}) that as   ${\vec r} \rightarrow \infty$, 
\beq a_{\sigma} \rightarrow b_{\sigma} e^{i\phi(\theta)} \label{asym}\eeq
where $b_{\sigma}$ is some constant spinor . The important point is that
 all four components of $a_{\sigma}$ have the \underline{same} asymptotic
phase $\phi$ which may depend on the spatial angle $\theta$. The underlying
reason is that the system is invariant under the same single U(1) gauge 
transformation  \ref{GT} acting on all the four components of $a_{\sigma}$.
Finally, the phase function $e^{i\phi(\theta)}$ as
 ${\vec r} \rightarrow \infty$
is a mapping of one circle (spatial infinity) into another (the U(1)
group manifold), and can therefore be divided into homotopy classes
characterised by a winding number 
\beq Q \ = \ -{i \over 2\pi} \int d^{2}r \bigg[  \epsilon_{\mu\nu} 
(D_{\mu} a_{\sigma}) ^{*} \ (D_{\mu} a_{\sigma}) \bigg] \label{Q} \eeq
For more details supporting these results see ref \cite{Raj}.
Exact soliton solutions for the minimal $CP_{3}$ system 
also known analytically
in terms of analytic functions. Those will not however hold for our full
system \ref{INTE} which has to be used when the layer 
separation $d \neq 0$. The solutions will have to be obtained numerically
by using appropriate ansatz. But the boundary condition \ref{asym} and 
the winding number  classification will still hold. They can be used to decide
what forms of intertwined spin-pseudospin solitons are permitted in
double layers.

An important consequence of the common phase boundary condition \ref{asym}
is that certain spin textures one can imagine having   for two separate  
single layers are
 not  permissible in the double layer system.
Consider a single layer at $\nu = 1$ carrying a Skyrmion with 
winding number $n$.
This is a finite energy configuration which 
can be described  by a 2-component spinor, say,
\[ \left(  \begin{array}{c}
	\lambda (r) \\
        f(r)e^{in\theta} 
                \end{array}\right)  \]
obeying boundary conditions as  $ r  \rightarrow \infty $ given by  
\beqarr 
    \lambda(r)   & \rightarrow   & 0 \nonumber \\
    f(r)     & \rightarrow  & 1 \nonumber \eeqarr
and as  $ r  \rightarrow 0 $
\beqarr 
    \lambda(r)   & \rightarrow   1 \nonumber \\
    f(r)     & \rightarrow   0       \label{BC3} \eeqarr 

One can have two such layers, widely separated,
 with two different spin-winding numbers $n$ and 
$m$ respectively. Nothing prohibits this .
 However suppose the two layers are part of a $\nu = 1$ double layer system
 at finite $d$ , and are described by a $CP_{3}$ 4-spinor
\beq {1 \over \sqrt{2}} \pmatrix{ 
	\lambda_{1} (r) \cr
        f_{1}(r)e^{im\theta} \cr
        \lambda_{2}(r) \cr
         f_{2}(r)e^{in\theta} 
            \cr} \label{nogood} \eeq
This would violate the condition \ref{asym} since asymptotically the 
second and fourth components would have different phase functions. Such
a texture is forbidden as per our analysis and indeed if one calculates
its energy by inserting it in \ref{INTE} one will find the energy diverging 
logarithmically. The divergence  comes from the angular derivative of 
${1 \over r^2} \partial_{\theta}^{2}$ contained in the Laplacians 
$\nabla^2$ in \ref{INTE}. That yields a contribution to the energy density 
proportional to         
\beq \frac{n^{2} + m^{2}}{2r^{2}} -\frac{(n+m)^{2}} {4r^{2}} \label{div} \eeq
as $ r  \rightarrow \infty$ which will lead to a logarithmic divergence
unless n = m. 
At the theoretical level the reason for this can be traced to the 
reduction of gauge symmetry discussed earlier, from $U(1) \times U(1)$ 
to $U(1)$ when two  layers are  together.

Keeping in mind this constraint of equal spin-winding numbers in each layer,
let us illustrate  non-trivially intertwined spin-pseudospin configurations
with the following example that \underline{is} allowed :

\beq   A \pmatrix{  
               \lambda_{1} \cr
         z-b \cr
	\lambda_{2} \cr
            z+b   \cr} \label{exact} \eeq
Here $\lambda_{1,2}$  and $b$ are non-zero constants while $z$ is the 
complex coordinate on the plane. $ A = 
\big(\lambda_{1}^{2} + \mid z-b \mid^2 + \lambda_{2}^{2}
 +   \mid z+b \mid^2 \big)^{-1/2}$ is the normalisation factor. 
Asymptotically, the first and third components  of \ref{exact} both 
behave as ${1 \over \sqrt{2}}
  e^{i\theta}$ while the other two components vanish. This is therefore a
  permitted (energetically finite) $CP_{3}$ configuration with winding
  number $Q = 1$. 
  
  One can see that this example is so designed that within each layer 
  the spin texture looks like that of a single 
Skyrmion, while at the same time it is also a "bimeron" in the "psuedospin
of  the down-spin 
component" (contained in the second and fourth components of the 4-spinor of
\ref{exact}. See ref (\cite{Ghosh2}) for more on bimerons.) 
But, we should remember that the the upper and lower layers are
not separately normalised in the example \ref{exact}. As ${\vec r}$ varies 
so does the relative
charge density in the two layers. Thus the spin vector in the upper (or lower)
 layer in \ref{exact} will not be a unit vector at every point unless it is
 is locally renormalised by the charge density of that layer at that point. 
 Similarly, 
 while the pseudospin of the down-spin component in the example \ref{exact}
   forms a  bimeron, this pseudospin will also be a unit vector 
   at each ${\vec r}$ only after being renormalised by the down-spin density, 
   which varies from point to point. Such renormalisation can be
   achieved by writing any general
   $CP_{3}$ 4-spinor (\ref{spinor}) in terms of spin and pseudospin 
polar angles.
\beq  a_{\sigma} \ = \ \pmatrix{
	\cos \frac{\alpha}{2} \cos \frac{\theta_{u}}{2} \cr
        \cos \frac{\alpha}{2} \sin \frac{\theta_{u}}
        {2}e^{i\phi_{u}}   \cr 
        \sin \frac{\alpha}{2} \cos \frac{\theta_{l}}{2}
         e^{i\beta}   \cr
        \sin \frac{\alpha}{2} \sin \frac{\theta_{l}}{2} 
         e^{i(\beta+\phi_{l})}
            \cr} \label{angles} \eeq
where the angles 
$\theta_{u.l}$, and  $\phi_{u,l}$ are the polar angles of the spin in the
upper(lower) layer while $\alpha$ and $\beta$ are the polar angles of the
pseudospin, each of which is a 
the function of the coordinate ${\vec r}$. (Recall that  $CP_{3}$ spinor 
has 6 real gauge invariant degrees of freedom. ) Suppose we   tentatively 
define, using these polar angles
 the familiar expression for the  spin-Skyrmion number in each layer by
\beq  n_{u,l}  \ =  \ {1 \over 4\pi}\int d^{2}r \epsilon_{\mu \nu} 
 \partial_{\mu} 
( cos \theta_{u,l}) \ \partial_{\nu} (\phi_{u,l}) \label{Qs}\eeq
One can then verify that the configuration \ref{exact} indeed yields unit
 spin-winding numbers $n_{u,l} =1 $ in each layer.
 
 Similarly , to get the pseudospin winding number one uses an
 alternate  parametrisation  of the same 4-spinor.

 \beq a_{\sigma} \ = \ \pmatrix 
	{\cos \frac{\alpha_{\uparrow}}{2} \cos \frac{\theta_{s}}{2} \cr
        \cos \frac{\alpha_{\downarrow}}{2} \sin \frac{\theta_{s}}
        {2}e^{i\phi_{s}}   \cr 
        \sin \frac{\alpha_{\uparrow}}{2} \cos \frac{\theta_{s}}{2}
         e^{i\beta_{\uparrow}}   \cr
        \sin \frac{\alpha_{\downarrow}}{2} \sin \frac{\theta_{s}}{2} 
         e^{i(\beta_{\downarrow}+\phi_{s})} \cr} \label{angles'} \eeq
Then the pseudospin winding number for the down spin component, for example,
can be written as 
  \beq n_{ps}(\downarrow) \ = \ {1 \over 4\pi}\int d^{2}r \epsilon_{\mu \nu} 
  \partial_{\mu} (cos \alpha_{\downarrow}) \ \partial_{\nu}
   (\beta_{\downarrow}) \label{Qpseudo}\eeq
Again, the example \ref{exact} happens to yield $n_{ps}(\downarrow)
 = 1$ in addition to, as we have seen, $n_{u,l} = 1$ .
Thus the example \ref{exact} illustrates an intertwined spin-
pseudospin topological configuration, containing the spin texture of a 
 Skyrmion in each layer and the pseudospin texture of 
a bimeron in the down-spin component.

One should however be cautioned that 
 there is only one  true topological charge $Q$ in the full $CP_{3}$ theory,
   given in eq. (\ref{Q}).
   Although the above example (\ref{exact})  contains the texture
 of two Skyrmions 
and a bimeron,  its $CP_{3}$  topological index $Q$  obtained by
 inserting it into eq. (\ref{Q}) will come out to be
 not 3 , but unity. 
 The separate sub-charges for spin and pseudospin defined in (\ref{Qs}) and 
(\ref{Qpseudo}) in  general do not have the same sanctity as they would 
 have had 
 for Skyrmions in a single layer or bimeron in a spinless problem. 
 Although in the above example these separate spin and pseudospin 
 winding numbers turn out to be integers, in general
 they need  \underline{not}  be integers , or more importantly ,
  be conserved in time. They are not protected by 
  homotopy considerations in our full 4-component theory. The angles 
 $\theta_{u,l} $ etc used in \ref{angles} cannot always be
obtained from the original components $a_{\sigma}$ of the 
4-spinor \ref{spinor}
, since they 
are not defined at those singular  points where $\alpha = \pi , 0$
  respectively. A similar remark holds for the other angles used above.
  The numbers $n_{u,l}$ and $n_{ps}$ can change 
  in  time due to leakages through such singular points .
   It is however interesting to note that the
  exact $CP_{3}$ winding number can be rewritten in expanded form 
  using the angles defined in \ref{angles} into parts that can 
  be attributed to winding of spin and pseudospins. This also brings out the
intertwining of spin - psedospin texture . We have, 
\beqarr 
Q &=& \frac{1}{4\pi} \int d{\bf r} \epsilon^{\mu \nu} \bigg( \partial_{\mu}
(cos \alpha) \big{[} 
\big{(}  \frac{1}{2}(1- \cos \theta_{u})\partial_{\nu} \phi_{u} - 
\frac{1}{2}(1 - \cos \theta_{l})
\partial_{\nu} \phi_{l} -\partial_{\nu} \beta \big{] }
\nonumber \\  
  & &
- F_{u}(\vec r)\partial_{\mu}(\cos \theta_{u})\partial_{\nu} \phi_{u} 
- F_{l}(\vec r)\partial_{\mu}(\cos \theta_{l})\partial_{\nu} \phi_{l} \big{]}
\bigg)\label{CFIL} \eeqarr
 where $F_{u}(\vec r) = (1/2) (1 + \cos \alpha(\vec r))$ and
$F_{l}(\vec r) = (1/2) (1 - \cos \alpha(\vec r))$ are respectively 
the same quantities as in (\ref{F1},\ref{F2})  and 
denote the number density in the top and bottom layers.
For the spinless (spins fully frozen) case one can set $\theta_{u}
,\theta_{l} , \phi_{u} and
 \phi_{l} $  to be constants. Then 
 one will recover the pseudospin topological charge
formula 
\beq n_{ps} = -\frac{1}{4 \pi} \int d{\bf r} \epsilon^{\mu \nu} \partial_{\mu}
(\cos \alpha) \partial_{\nu} \beta. \eeq
Similarly for a single layer (say the upper layer) case one can set
$\alpha = 0$ and recover
the spin winding number formula 
\beq n_{u} = -\frac{1}{4 \pi} \int d{\bf r} \epsilon^{\mu \nu} \partial_{\mu}
(\cos \theta_{u}) \partial_{\nu} \phi_{u}. \eeq
In the general, where both spin and pseudospin havesome intertwining texture,
the full topological charge will receive contributions from the windings 
of all these, as given in \ref{CFIL}. 

Finally, the very simple example \ref{exact} not only illustrates a nontrivial
intertwining texture , it is also an exact solution of the prototype $CP_{3}$
theory \ref{proto}
 since its components are analytic functions (see ref \cite{Raj}).
But, for our full theory in the presence of a non-zero layer
separation and with Zeeman and tunnelling energies ,
classical  solutions minimise the full energy functional  \ref{INTE}, 
 have to be obtained numerically by solving the 
non-linear coupled partial differential 
equations that the minimisation conditions yield. The simple analytical
example \ref{exact} however will guide us in setting up the desired
ansatz for the numerical solution with appropriate boundary conditions
 so that  intertwined textures which are nontrivially wound in both spin
 and pseudospin can be obtained for our full theory. An illustrative
 family of such solutions is obtained in the next section.

\section{FIELD EQUATIONS AND THEIR SOLUTIONS}
Classical solutions that minimise the full energy functional \ref{INTE}
 have to be obtained numerically. To do this we use the parametrisation
 of the spinor components of the form
 
\beq  a_{\sigma} \ = \ \pmatrix{
	\cos \frac{\alpha}{2} \cos \frac{\theta_{u}}{2} \cr
        \cos \frac{\alpha}{2} \sin \frac{\theta_{u}}
        {2}e^{i\phi_{u}}   \cr 
       0   \cr
        \sin \frac{\alpha}{2} 
         e^{i \phi_{l}}
            \cr} \label{ansatz} \eeq

One can see that this is a sub-family of the general case in eq (\ref{angles})
where for simplicity we have set  $\theta_{l}$  equal
to $\pi$ and  $\beta = 0$.

We will look for numerical solutions which
 would have corresponded ,if  the energy had been of the simple prototype
 functional (\ref{exact}), to its exact analytic solution

\beq   A \pmatrix{  
               \lambda \cr
         z-b \cr
	0 \cr
            z+b   \cr} \label{sub} \eeq

 This configuration represents a spin skyrmion in the upper layer intertwined 
with a bimeron in the " pseudopsin of the downspin 
component". It does not have any non-trivial winding in the real spin of the
lower layer (though the fourth component in the spinor
 varies over the coordinate space the spin will be 
always down). To ensure that our numerical solutions have the
same topological properties as well as similar profiles as
  the prototype spinor (\ref{sub}) we impose
the same boundary conditions on the components of the former as obtained in
the latter, both 
 asymptotically and at the meron centers 
  centres $x = \pm b$.

In terms of the ansatz \ref{ansatz}
 the local energy functional (\ref{INTE}) takes on the form

\beqarr E_{C}  & = & \beta_{m}  \int d \vec r \cos^{2}\alpha 
  +2\rho^{s} \int d \vec r \bigg{[} \frac{1}{4} \bigg{(} 
(\vec \nabla \alpha)^{2}
+ \cos^{2}\frac{\alpha}{2}(\vec \nabla \theta_{u})^{2} \bigg{)}
\nonumber \\    &   & \mbox{} +  \frac{1}{4}
\bigg{(} (1 +\cos \alpha)(1 - \cos \theta_{u})(
\vec \nabla  \phi_{u})^{2} +
2(1 - \cos \alpha)(\vec \nabla \phi_{l})^{2}  \bigg{)} \nonumber \\
                &   & \mbox{} - \frac{1}{16} \bigg{(} (1 +\cos \alpha)^{2}
(1 - \cos \theta_{u})^{2}(\vec \nabla \phi_{u})^{2} +
4(1 - \cos \alpha)^{2}(\vec \nabla \phi_{l})^{2} \nonumber \\
                &   & \mbox{} + 4(1-\cos^{2} \alpha )
(1 - \cos\theta_{u})
\vec \nabla \phi_{u} \cdot \vec \nabla \phi_{l}  \bigg{)}
\bigg{]} \nonumber \\ 
		&   & \mbox{} +(\rho^{s} - \rho^{l})  
\int d \vec r\bigg{[} \frac{1}{2}(
(\vec \nabla \cos \alpha )^{2} - (\vec \nabla \alpha)^{2} ) \nonumber \\
&   & \mbox {} +\sin ^{2} \alpha \bigg{(} -\frac{1}{8} 
(\vec \nabla \theta_{1})^{2}
 -\frac{1}{4} \bigg{(} (1 - \cos \theta_{u})
(\vec \nabla \phi_{u})^{2} +2
(\vec \nabla \phi_{l})^{2}  \nonumber \\
&   &  \mbox{} + 2(1 - \cos \theta_{u})
 \vec \nabla \phi_{u} \cdot \vec \nabla \phi_{l} \bigg{)} \bigg{)}
\bigg{]} \label{Ebs}\eeqarr  

This energy functional has to be minimised with respect to all the 
angle fields in the anstatz.
As we did in our earlier work on the spinless problem \cite{Ghosh2},
here too we will use  the 
bipolar co-ordinate system \cite{Margenau} to describe the spatial plane.

\beq \eta \  = \ ln |z-a| - ln |z+a|  \ \ ; \ \ \phi \ = \ arg(z-a) - arg(z+a)
\eeq 

We have already elaborated in \cite{Ghosh2} 
  the advantages of this co-ordinate
system  when one has to impose the bimeron
-type  boundary conditions. However here the advantages of introducing such an
unfamiliar co-ordinate system is not as much  
as in the simple spinless bilayer problem of \cite{Ghosh2}  because 
the  ansatz here is not symmetric between the two layers.
 Consequently unlike the spinless case  
$\cos \alpha$ is no more antisymmetric about $\eta$ = 0 axis.
 All these features 
along with the fact that the energy minimisation
unavoidably requires solving coupled non-linear partial diffential equations
( p.d.e.) render the numerical exercise much more  complicated here . What
 we have done under these circumstances is the following. 

We have solved the field equations
numerically for  the case  where just
 the capacitance term is added to the minimal $CP_3$ energy . 
>From our earlier calculations 
we know that this term is going to change the  
solutions considerably. The terms in  each equation with the coefficient
$(\rho^{s} - \rho^{l})$,  which accounts for the anisotropy in 
the exchange energy is not included in the process of numerical 
integration.
 As a justification of such simplification 
we can say that the anisotropic terms which involve higher order gradients 
of the spin pseudospin field will have less pronounced effect compared 
to the capacitance term on the  solutions.
This has been graphically shown in Fig. 1 
and 2 of our earlier work \cite{Ghosh2}.
Even after this drastic  simplification  we are still left with  
solving four coupled 
non-linear p.d.e's . For example, the 
equation  which is obtained by extremising the energy w.r.t 
$\cos \alpha$ is

\beqarr 
\bigg(\frac{\delta E_{C}}{\delta \ca}\bigg)_{\rho^{s} = \rho^{l}} 
& = & 2 \beta_{m}  Q_{s}^{2} \ca 
 + \rho^{s} \bigg{[} -2\frac{ \ca (\vec \nabla \ca)^{2}}{(1-
\cos^{2} \alpha)^{2}} \nonumber \\
&   &\mbox{}  - 2\frac{\nabla^{2} \ca}{1 - \cos^{2}\alpha} 
+ (1 - \ct )(\vec \nabla \phi_{u})^{2} -2
(\vec \nabla \phi_{l})^{2} \nonumber \\
&   &\mbox{} - \frac{1}{4}  (1 + \ca)(1 - \ct)^{2}
(\vec \nabla \phi_{u})^{2} \nonumber \\
&   &\mbox{}  + (1 -\ca)(\vec \nabla \phi_{l})^{2} +
 \ca(1-\ct)(\vec \nabla \phi_{u} \cdot \vec \nabla
\phi_{l})  \bigg{]}  \nonumber \\
& = & 0 \label{calpha} \eeqarr

where
 \beq Q_{s}^{2} \ (\eta, \phi) \ =  
\frac{b^{2}}{({\cosh{\eta}-\cos{\phi}})^{2}}
\label{Evar} \eeq
is the Jacobian of this coordinate transformation
and all gradient operators are defined in the bipolar-coordinate system
\cite{Margenau}. Similarly we wil have three more equations  obtained 
by extremising the  energy functional
with respect to $\theta_{u} , \phi_{u} , \phi_{l}$ and then writing the
resulting equations in bipolar co-ordinates. We will not display them here. \\

\subsection{Numerical Procedure}
The numerical procedure is almost same as that in  Ref. \cite{Ghosh2}.
Here also one can see that the Jacobian factor $Q_{s}$ in the 
first term of the eq. \ref{calpha}  
 is singular at the point 
$(\eta=0,\phi=0)$. The behaviour of $\cos \alpha$ near this
point is also going to be 
same as the  behaviour of $m_z$ in ref. (\cite{Ghosh2}).
 The major difference compared to the earlier problem however 
comes from the fact that it is no longer  
sufficient to find out the solutions in one quadrant and  
get the rest from symmetry considerations. 
We have to solve this problem 
on both side of the $\eta = 0$ axis since our starting ansatz solution
is not completely antisymmetric
around $\eta = 0$.
During the numerical work one also has to be careful about 
the different branches of the angles $\phi_{u,l}$. As one needs to
 integrate the
equations on the both sides of the $\eta = 0$ axis the size of the mesh
on which we have to discretise the field equations becomes larger compared
to the earlier case of spin-frozen double layer problem \cite{Ghosh2}.
Also here we have to solve four coupled p.d.e's simultaneously. 
 This simultaneous increase in the number of lattice points
as well as independent fields  demands that we have to invert a 
huge determinant in the Newton-Raphson procedure \cite{numer}
while improving over the initial guess solution. This forces us
to increase the lattice constants of the mesh slightly compared to 
what we have done in our earlier work \cite{Ghosh2}. 
However we have checked that the error
introduced in this way is not very high. In the next 
subsection we shall present
our results along with the discussion.

\subsection{Results and Discussion}
Our solutions of eq.(\ref{calpha}) along with the other three
field equations 
yield the spatial behaviour of the $CP_{3}$  fields
parametrised in term of the angles $\alpha$ , $\beta$ ,$\theta_{u,l}$
and $\phi_{u,l}$. The calculations are done iteratively. We start with
 the simple analytical spinor \ref{sub} which is the exact solution 
 when the capacitance and anisotropy terms in the energy (\ref{Evar})are 
 absent. Then the capacitance  term in the equation is
 introduced in small steps and the corresponding solution obtained
 numerically.
We have performed  several calculations each starting from
different initial values of the constants
 $\lambda$ and b . The constant
 $\lambda$  represents the starting value of the first component
 of the spinor in the iteration process. It stands for the spin-Skyrmion 
 size  in the $CP_3$ limit, but when subsequent iterations are performed
 in the presence of other energy terms, it is replaced by a space
 dependent solution. But the parameter $b$ is fixed for a given
  calculational run. It represents the meron separation and enters into the 
 equation (\ref{calpha}) 
explicitly through the first (capacitance) term. While we do calculations
for different values of $b$, the optimal value of $b$ will
have to be obtained by minimising the energy with repsect to it.
We will return to this point later.

We  present below  the salient features of our numerical results.
The major feature we want our  numerical solution  to have
is  the intertwining of the spin skyrmion with the pseudopsin
bimeron . We would also like to show the leakage of electrons of either spin
from one layer to another as we move in space, 
as a fallout of this intertwining. To show this 
, we have plotted both $\cos \alpha$ as well as $\cos 
\alpha_{\downarrow}$ as a function of $\eta$ for a set of values 
of the angle $\phi$ in Fig.1 and Fig. 2 respectively.
These solutions correspond to layer separation $ d= 0.6l$ ,
bimeron center-separation $b = 2.5l$ and starts from an initial value
of  $\lambda = 1$ in the starting trial solution \ref{sub}. 
The sequence of curves shown
correspond to $\phi$ equal to 0.09$\pi$, 0.36$\pi$, 0.63$\pi$, and 0.90$\pi$
respectively with the outermost one belonging to $\phi$ equal to 0.09$\pi$.
 As we have discussed earlier \cite{Ghosh2}, 
Note from the definition of the bipolar coordintaes
that spatial infinity in x-y plane corresponds to  $\eta$ and $\phi$ both 
equal to zero. As we approach this point in the $\phi , \eta$ plane, 
the solution
should damp exponentially as
$exp(-\frac{\kappa}{\sqrt{\eta^{2}+\phi^{2}}})$ where 

\beq \kappa \ = \ \sqrt{\frac{2\beta}{\rho_{A}}} b \eeq.
Correspondingly we see in
Fig. 1 and 2 that the low $\phi$ curves rise very 
slowly as $\eta$ increases away
from zero. 

The  interesting point to note about these solutions is that in Fig. 1 
$\cos \alpha$ approaches different (absolute) asymptotic values as $\eta$
approaches $\pm \infty$.(These are respectively the centers of the two 
merons that form the bimeron. Although computational limitations allow us 
to go only upto values 0f $\eta = \pm 3$, it is clear from the figure that
asymptotic behavior has been obtained ). 
 This asymptotic behaviour is extracted directly
 from the analytic ansatz \ref{sub} and implies the leakage from the
pseudospin to spin. It is useful to remember at this point that
we have a bimeron only in the "pseudospin of the down-spin componenet"
whereas  $\cos \alpha$ represent the z-component of the total pseudospin.
This is realised in Fig. 2. Here $\cos \alpha_{\downarrow}
(down-spin)$ represent 
the z-componenet of the "pseudospin of the down-spin component".
It is completely antisymmetric about $\eta = 0$ and approaches 
$\pm 1$ as $\eta$ approaches $\pm \infty$. This behaviour
is same as the behaviour of $m_z$ in the spinless bilayer case.
This is how we can  extract from our results 
the pure bimeron by suitably partitioning the pseudospin
into different spin components.  

Since bipolar co-ordinates are not  very familiar
 we have given an alternate representation of the above 
results through a vector-plot in the physical x-y space in Fig. 3 and 4. The
values of the parameters in these figures are same as those in 
Fig.1. In Fig. 3 the magnitude of each arrow gives the 
absolute value of the transvere component of the total pseudospin
$\sin \alpha$ and it's angle with the x-axis gives $(\phi_u - \phi_l)$. 
One should note in this regard the  $(\phi_u - \phi_l) = \beta_{\downarrow}$
in the other parametrisation
 Here also one can see the length of the arrow does not quite vanish 
in one of the bimeron centres and makes this construction singular
at this point.
However when we look at the Fig. 4. where the magnitude of each 
arrow is $\sin \alpha_{downarrow}$ and the angle is again $(\phi_u - \phi_l)$
, the magnitude of the arrow vanishes at each bimeron centre and the
profile is no more singular. This makes Fig. 4  identical to the 
vector plot of the bimeron pseudopsin in the pure layer case \cite
{Brey}, \cite {Ghosh2}.

In Fig 5.we have given a similar vector plot for the spin skyrmion
in the upper layer.
 Here the  length of the each arrow corresponds to the  planar 
projection  of the spin in the upper layer ($\sin \theta_u$) and the 
its direction gives the azimuthal angle ($\phi_{u}$) of the projected vector
.This picture very clearly points out how the skyrmion winds in the 
azimutahl plane about it's centre at $x = b$
 Here also the layer separation $d$ and the starting values of $b$
and $\lambda$ are the same as those in  Fig. 1 .
This set of parameters  represents a typical example. 
 
Lastly, we have  evaluated the energy of these solutions for a
set of values of the meron separation parameter $b$ .The optimal value
of $b$ should be obtained by minimising the full energy as a function of $b$.
But if we include only the capacitance term and the (pseudo)spin-stiffness
 gradient terms in our energy, these will not lead to a non zero $b$, 
 i.e the textures 
will want to shrink to zero size. The reason is that under rescaling,
the capacitance term grows proportional to the square of the scale while
the gradient terms are scale invariant. Of course a change in $b$ will not
 result
 in just an overall rescaling of the solution. The shape of the solution
 will also change since $b$ occurs as a  constant multiplying the 
 capacitance term in the differential equation (4.5).  As a result,
 the $b$ dependence of the gradient and capacitance terms will not be simple,
 although qualitatively it should still drive the meron separation to zero 
 size. This can be seen in Table 1, where we show these energies for ten  
different values of $b$ for a fixed value of other parameters. 
 As expected the sum of both these energy contributions decreases 
 strongly with decreasing $b$, thus driving the texture to zero-size.
  In reality however, these two terms are just the first two
   terms in the gradient expansion of the full energy
   functional. Higher order terms in the 
   gradient expansion, if included, will make our nonlinear differential
    equation even more difficult to solve, but they can 
    offset this tendency to shrink.

   In particular, one prominent higher gradient contribution to the energy is
 the Coulomb interaction between different portions of the topological
  charge densities. It is given by (see\cite{Arov}) :
\beq E_{Coul}  \ = \ \frac{1}{2}\int d{\vec r}d{\vec r'}
V({\vec r}-{\vec r'})
\delta \rho({\vec r}) \delta \rho({\vec r'}) \label{3COL} \eeq
where $\delta \rho({\vec r})$ is the $CP_3$ topological charge 
density given by the integrand of the r.h.s of the equation (\ref{Q}).

Inclusion of the contribution of this term into our 
differential equation for the
texture will introduce a  $\it{nonlocal}$ nonlinear term, which will make 
it very difficult for us to solve it numerically. We can however make the 
following estimate. We can insert our texture solution , obtained without
 the Coulomb term , into the Coulomb energy integral and evaluate it as a
 function of $b$. This contribution is also shown in the Table 1. 
As expected, the coulomb repulsion 
energy $E_{Coul}$ decreases with increasing $b$. This 
term would like to keep the merons farther apart, and 
  offset the  tendency
 to shrink because of the other terms. 
 Thus one may hope to  get an optimal bimeron separation at which
the sum of all these three energy contributions will get minimised as a 
function of $b$.

  In  Table 1  we have presented our calculation  for the layer 
separation $d = .8l$ where  different contributions to the total energy
are shown along with the their sum, the "total energy".  For this particular 
layer separation a distinct minimum is obtained around $b = 2.0 l$.
 
To see whether this behaviour is common to other layer separations
 we have plotted in fig.6 the total energy 
$E_{total}$ as a fuction of bimeron
separation $b$ for a set of layer separations. 
The three sets of points in this figure corespond to three
different layer separations, namely $d$ equal to $.5l$, $.7l$ and $.8l$.
All these three curves show distict minima for the total energy as a 
function of bimeron separation. We have 
 also provided in Table 2 the size ($b$) and total energy of the optimal
bimeron for five values of layer separation.  

Notice from Table 2 that the optimal
meron separation $b$ decreases with the increase of layer separation $d$
. For $d=.8l$ 
the optimal separation 
is around $b=2.0l$ and gradually increases to $ b=2.5 l$ 
for $d=0.4l$ .  For the 
case of pure layer bimeron Brey et. al. also found (\cite{Brey})
 a similar behaviour.  
This behavior could  be attributed to the following. 
By lowering the layer separation (decreasing the $\frac{d}{l}$ ratio) one  
increases the relative importance of Coulomb repulsion 
among topological charge densities (coming from the intra-layer Coulomb 
energy)  to the Capacitance term (coming from the interlayer Coulomb 
repusion). Hence 
the balancing of the capacitance 
term by the Coulomb energy will take place at a larger  bimeron separation, 
thereby increasing  the size of the bimeron. 

 We found that  resulting energy of these solitons 
at their optimal sizes varies  very little as one changes the layer
separation. Although  bimerons of larger size at lower layer 
separations cost higher capacitance energy , the decrease in the 
Coulomb energy seems to fully offset  that. 
As a result the total energy remains 
almost the same for different  layer separations in the range $d=0.4 l$ 
 to $d= 0.8 l$ that we have studied.

An important question is whether our spin-pseudospin intertwined  solution
has a lower energy than other candidates among the low-lying
excitations. Prominent among these other low lying excitations
with whom such comparisons have to be done are (i) the 
particle-hole excitations and (ii) purely spin or pseudospin
textured solitons. To start with note that in the  minimal
prototype $CP_{3}$ system (valid in the $d=0$ limit ; see \ref{proto})
the energy is just equal to $ E_{CP} \ = \ 4\pi \rho^{s}Q  \ $ (see ref.
(\cite{Raj}). Now , a pure-spin skyrmion  in, say, one of
the layers can also be written in our $CP_3$ 4-spinor notation and
will  have a $CP_3$ topological number $Q =1$.   So
will a bimeron in pseudospin of some spin component. Therefore in the
prototype $CP_{3}$ system  our spin-pseudospin intertwined soliton with
 $Q=1$ will have the same energy as  a purely spin or pseudospin
textured soliton with $Q=1$. The intertwining will not cost more energy.
   All these energies which are equal,  
   are a quarter of that of a  particle-hole pair
(see (\cite{Arov}) which costs an energy of $\sqrt{{\pi \over
2}} \approx 1.25$ in units of ${e^2 \over \epsilon l}$ .
However the difference in the energies of these various types of
topological excitations come from the additional terms in the
full energy $E_{C}$ due to capacitance ,  anisotropy and Coulomb repulsion.
As we can see from the Table 2 that 
 our intertwined soliton over
a range of layer separation has energy  around .60 $\frac{e^{2}}{\epsilon 
\l}$. 
 It is encouraging 
that  a pair of these excitations would have somewhat lower energy than the
 particle-hole pair energy of 1.25. 

 Of course our computational accuracy is not very high, given that we are 
 limited in how many lattice points we can use. One must  also improve
  on the results by solving for the
  texture functions and their energy after including  
  single particle terms due to the Zeeman coupling
  and tunelling .  Examples of such  calculations can be found
in the case of $\nu =2$ by Pardes {\sl et. al.} \cite{Pard}
, but not for $\nu =1$ yet to date. Meanwhile, our result for
  the intertwined soliton at $\nu = 1$ and its energy
   at best raise hopes that they may be competetive as candidates for low 
   lying excitations in double layer systems with spin.

\vspace{1.0 cm}

{\bf Acknowledgement}\\
R.R. has benefitted greatly by discussions with Professor Allan MacDonald on 
spin-pseudospin systems. We thank Prof. R. Ramaswamy for allowing
us to use computational facilities in his lab. for some of the numerical work 
done here. The work of S.G. is partially supported by a
CSIR award no.9/263(225)/94-EMR-I .dt.2.9.1994.

\newpage
\bc
TABLE
\ec 
\noindent Table 1: Different contributions to the total energy ($E_{total}$)
 of spin-pseudospin intertwined 
solitons for a set of $b$ at a layer separation of $d = 0.8l$.
Here $E_{grad}$ refers to the gradient energy  
(isotorpic plus anisotropic), while
$E_{capa}$  is the capacitance 
energy and $E_{Coul}$ is the Coulomb interaction energy
between  topological charge densities . $E_{total}$ is the 
sum of these three contributions to the energy. The unit of 
energy is $e^2 \over \epsilon l$ and the unit of length is $l$.  

\begin{center}
\begin{tabular}{|c|c|c|c|c|}
\hline
$b$ &  $E_{capa}$   &   $E_{grad}$ & $E_{Coul}$ & $E_{total}$  
 \\
\hline
4.5  & 0.285 & 0.261 & 0.141 & 0.687  \\
\hline
4.0  & 0.250  & 0.251 & 0.152 & 0.653 \\
\hline
3.5  & 0.229  & 0.228 & 0.168 & 0.625 \\
\hline
3.0  & 0.205  & 0.223 &  0.183 & 0.611 \\
\hline
2.5   & 0.153 & 0.227 & 0.217 & 0.597 \\
\hline
2.3   & 0.143 & 0.227 & 0.225 & 0.595 \\
\hline 
2.0 & 0.126 & 0.203 & 0.262 & 0.591 \\
\hline 
1.8 & 0.121 & 0.192 & 0.290 & 0.603 \\
\hline
1.5 & 0.104 & 0.196 & 0.328 & 0.628 \\
\hline 
1.2 & 0.091 & 0.200 & 0.390 & 0.681 \\
\hline 
1.0 & 0.081 & 0.192 & 0.456 & 0.729 \\
\hline
\end{tabular}\\
\end{center}

\newpage
\bc
TABLE
\ec 
\noindent Table 2: 
The size (i.e.the optimal meron separation $b$) and the total energy
 ($E_{total}$) at different layer separations $d$. The unit
of energy is $e^2 \over \epsilon l$ and the unit of length is 
$l$.
\begin{center}
\begin{tabular}{|c|c|c|}
\hline
$d$ &  $b$   &  $E_{total}$  
 \\
\hline
0.8  & 2.0 & 0.59   \\
\hline
0.7  & 2.2  & 0.59  \\
\hline
0.6  & 2.3  & 0.60  \\
\hline
0.5  & 2.4  & 0.60 \\
\hline
0.4   & 2.5 & 0.59 \\
\hline
\end{tabular}\\
\end{center}

\begin{figure}
\label{fig1}
\caption{The solution  $\cos \alpha (\eta)$ of the field equations   
for a set of values for $\phi$.The curves correspond, as one  goes inwards, to
$\phi = 0.09\pi , 0.36\pi, 0.63\pi, 0.90\pi$  respectively with 
the outermost one corresponds to $\phi$ equal to 0.09$\pi$.
 The layer separation
 $d$ is equal to 0.6{\it l} and bimeron separation $b$ is equal to 
2.5{\it l}. The value of $\lambda$ in the analytic ansatz is $1l$ }
\end{figure}

\begin{figure}
\label{fig2}
\caption{The plot of  $\cos \alpha_{down spin(\downarrow)} (\eta)$   
for a set of values for $\phi$.The curves correspond, as one  goes inwards, to
$\phi = 0.09\pi , 0.36\pi, 0.63\pi, 0.90\pi$ respectively with 
the outermost one corresponds to $\phi$ equal to 0.09$\pi$. The 
layer separation
 $d$ is again equal to 0.6{\it l} and bimeron separation $b$ is equal to 
2.5{\it l}. The value of $\lambda$ in the analytic ansatz is also $1l$}
\end{figure}
\begin{figure}
\label{fig3}
\caption{This figure gives  the magnitude and direction 
of x-y projection of the total pseudospin 
  at different points on the plane. The magnitude of each arrow at a given 
point is 
$\sin \alpha$ and it's angle with the x-axis is $(\phi_{l} - \phi
_{u})$ at that point.
The layer separation and the  bimeron separation are same as in
Fig. 1 and 2}
\end{figure}
\begin{figure}
\label{fig4}
\caption{This figure gives  the magnitude and direction 
of x-y projection of the " pseudospin  in the down-spin component"
  at different points on the plane. The magnitude of each arrow at a given 
point is 
$\sin \alpha_{\downarrow}$ and it's angle with the x-axis is 
 $\beta_{\downarrow}$ at that point.
The layer separation and the  bimeron separation are same as in
Fig. 1 and 2}
\end{figure}
\begin{figure}
\label{fig5}
\caption{This figure gives  the magnitude and direction 
of x-y projection of the spin in the upper layer
  at different points on the plane.
At each point the magnitude of the arrow gives $\sin \theta_u$ and it's
direction with x-axis gives $\phi_u$ at that point.
The layer separation and the  bimeron separation  and initial
$\lambda$ are the same as in
the earlier figures}
\end{figure}
\begin{figure}
\label{fig6}
\caption{This figure gives  a plot of the total energy
E(total) as a function of the bimeron separation $b$
for three different layer separations, namely $d= 0.5l, 0.7l$ and $0.8l$
 The unit of energy is again $e^2 \over \epsilon l$}
\end{figure}

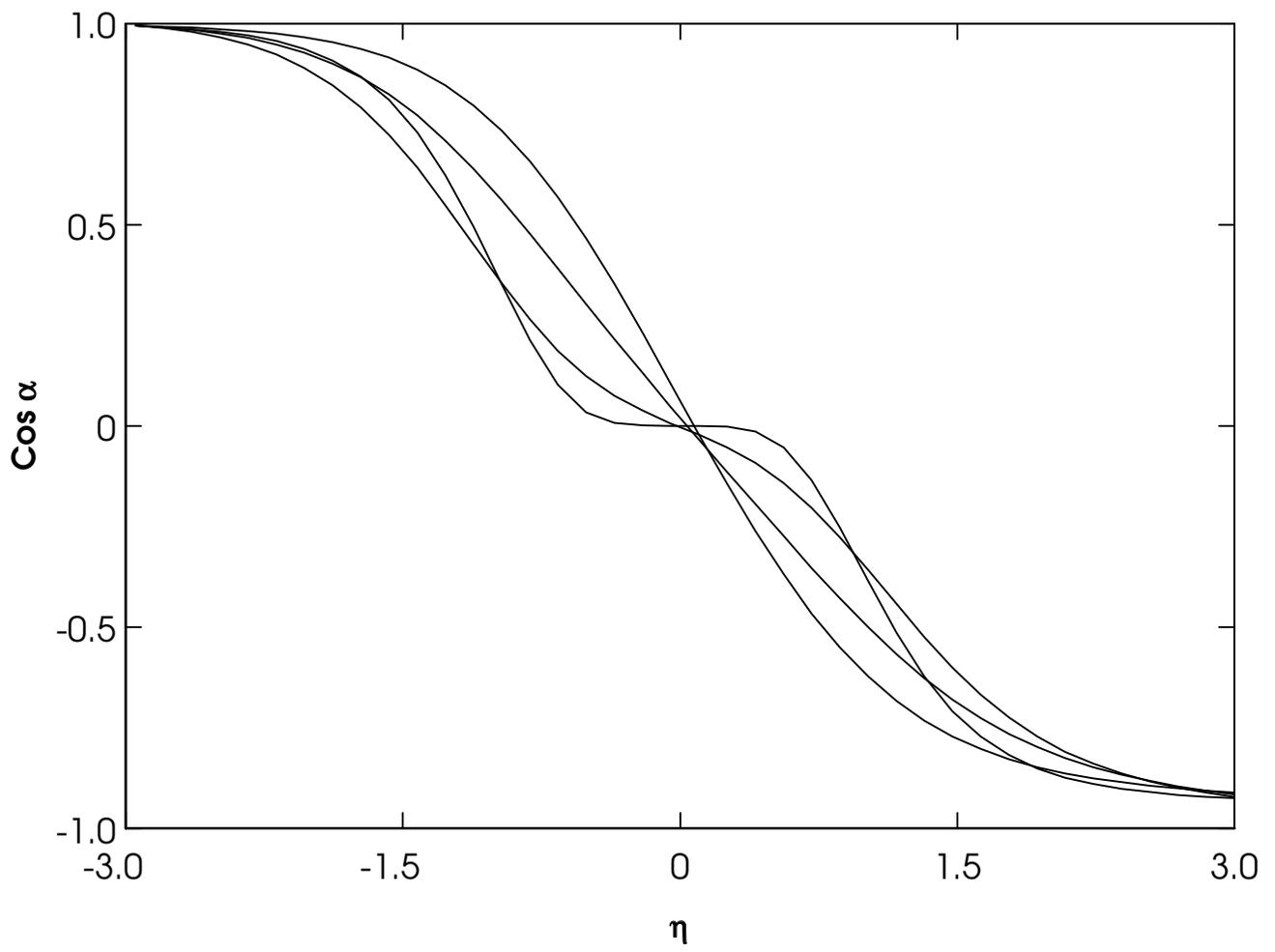
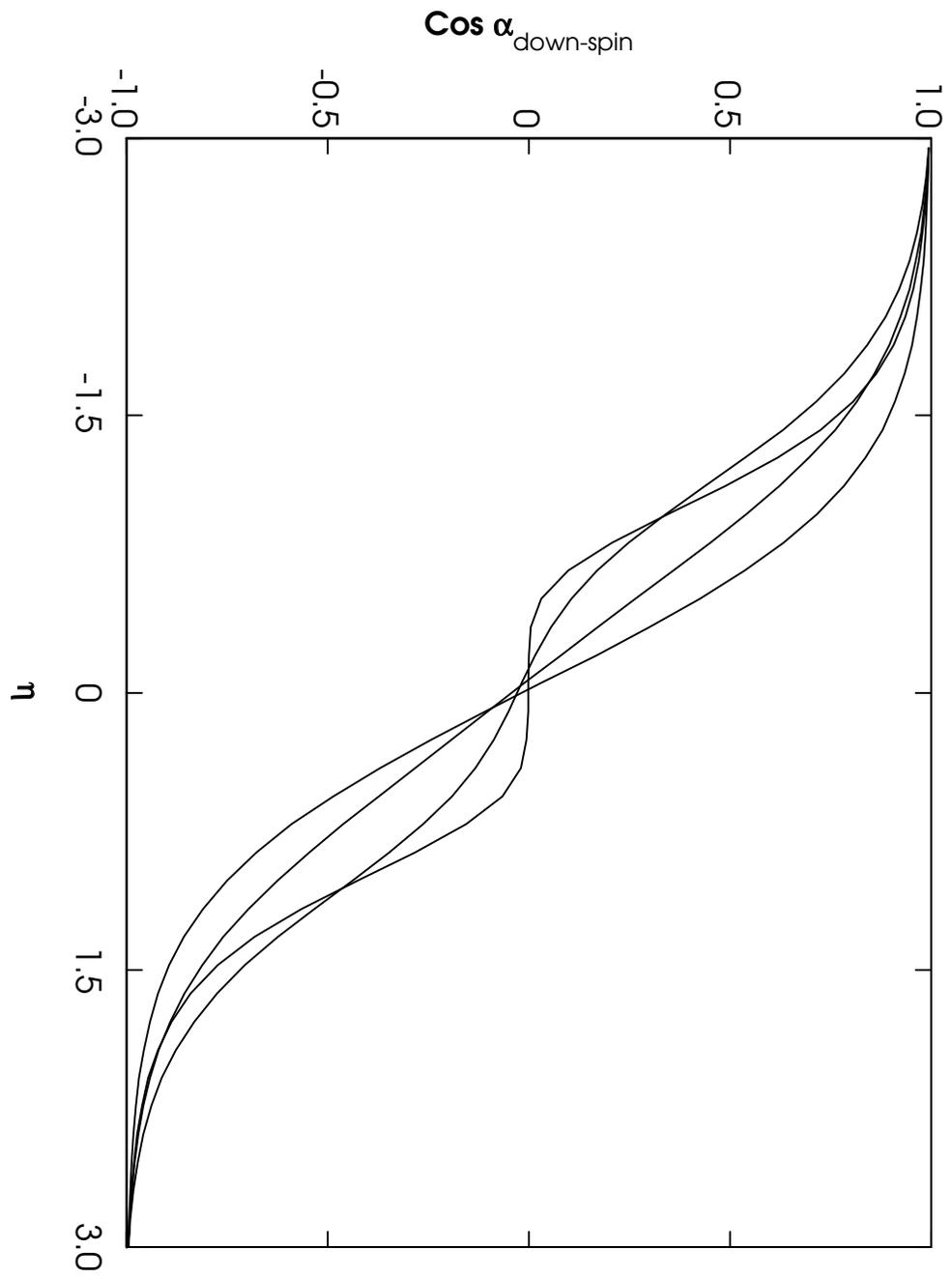
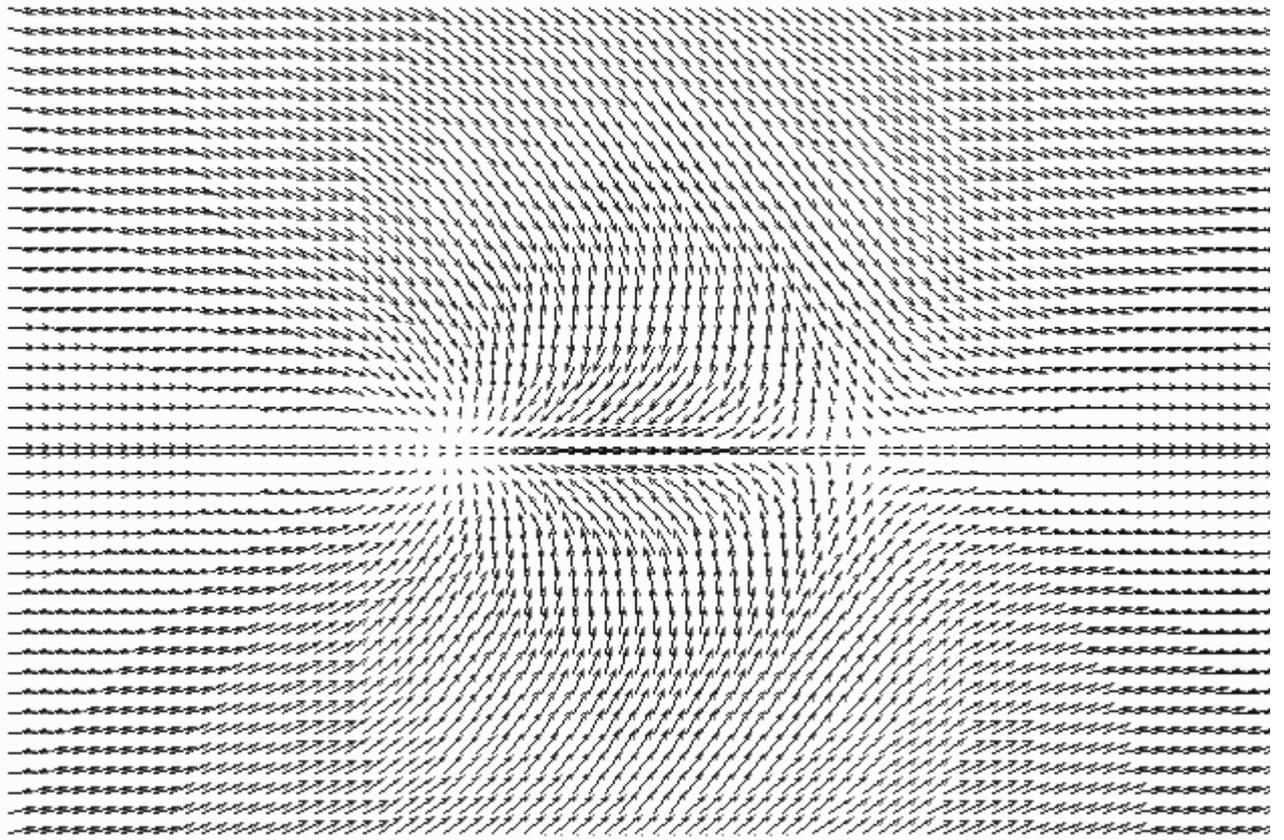
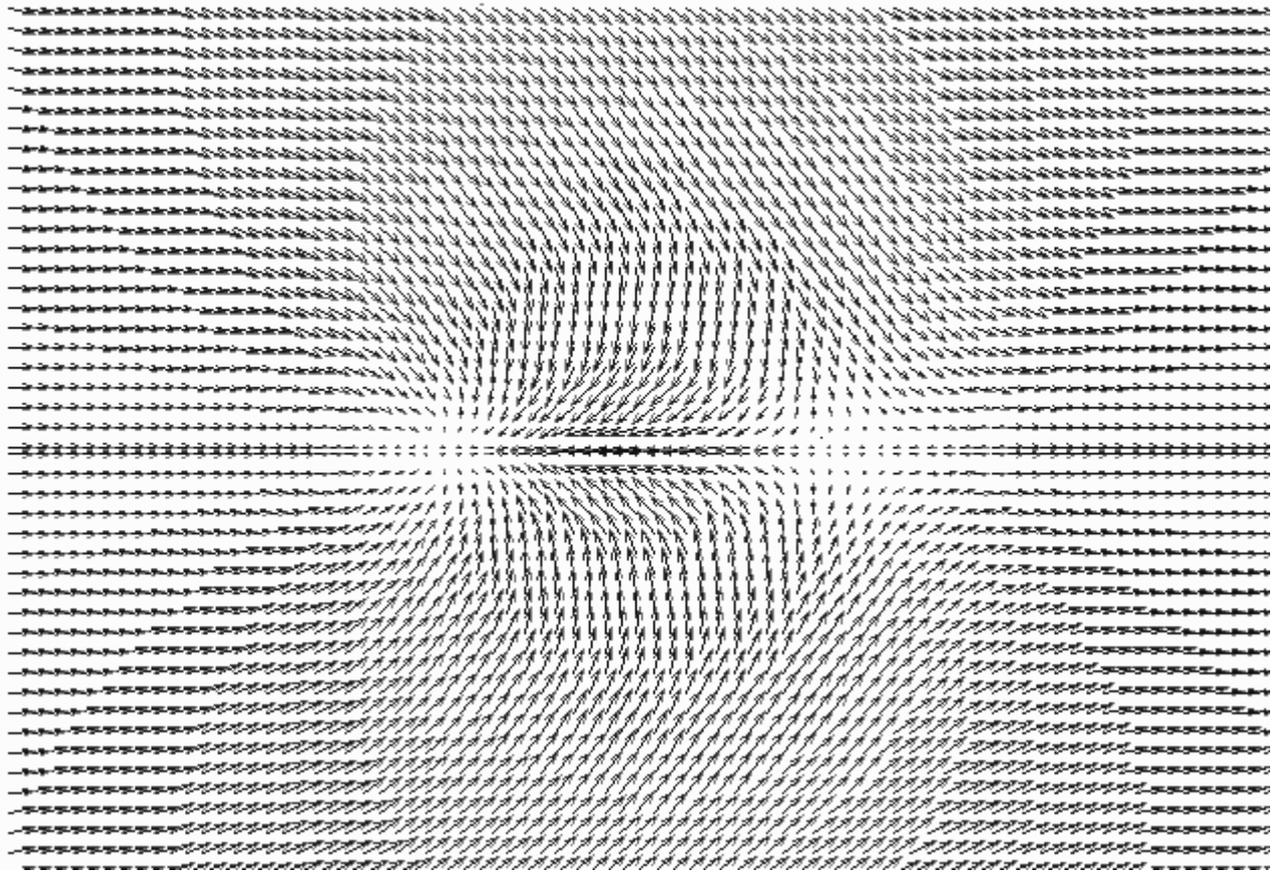
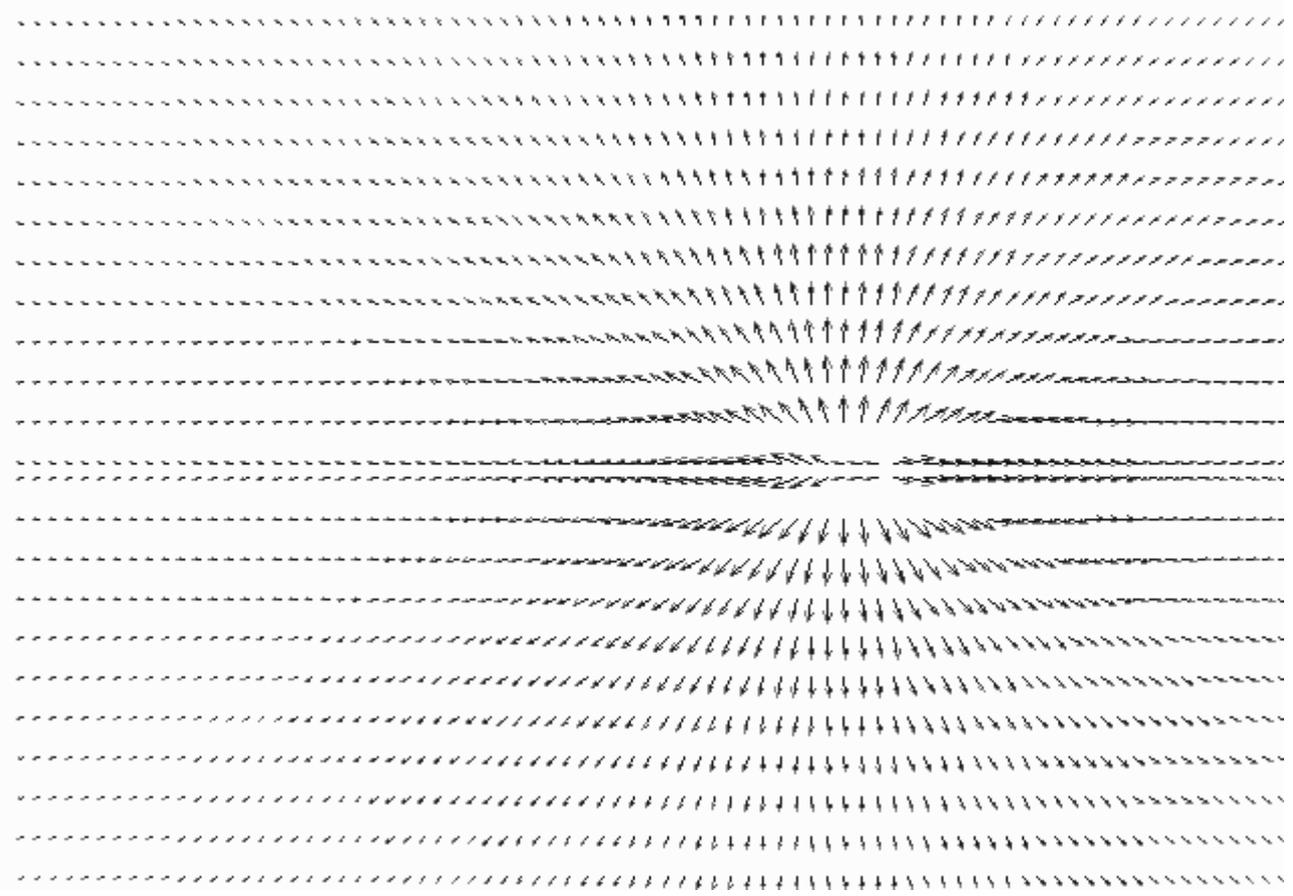
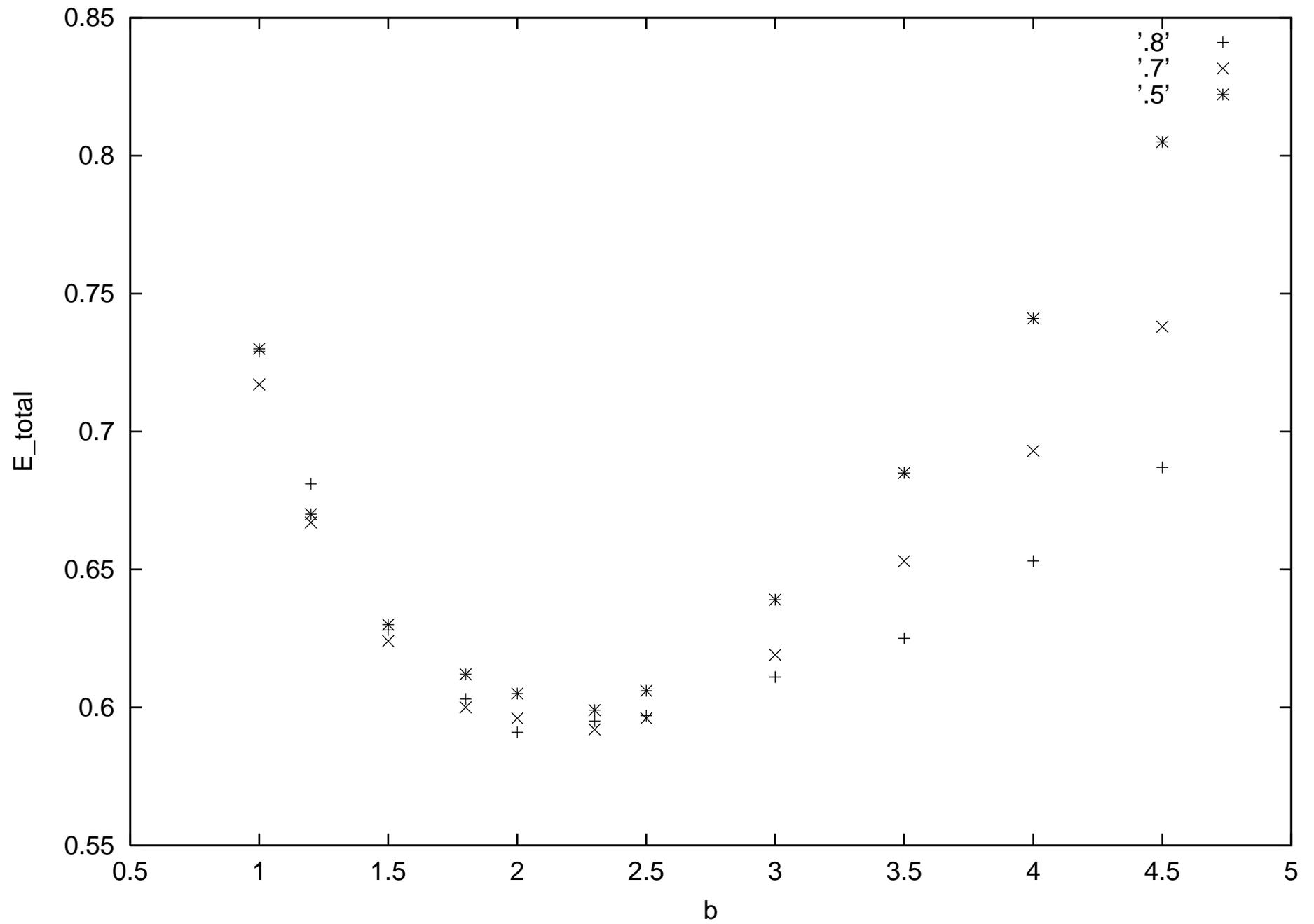
\end{document}